\documentclass[twocolumn]{aastex631}

\usepackage{natbib}
\usepackage{amsmath}
\usepackage{upgreek}

\begin{document}

\title{From Exo-Earth Detection to Atmospheric Characterization: Enabling Phase-Dependent Molecular Detection SNR in \texttt{EXOSIMS}}
\author[0000-0001-5477-8588]{Searra Foote}
\affiliation{Lunar \& Planetary Laboratory, University of Arizona, Tucson, AZ 85721, USA}
\affiliation{Habitability, Atmospheres, and Biosignatures Laboratory, University of Arizona, Tucson, AZ 85721, USA}

\author[0000-0002-3196-414X]{Tyler D. Robinson}
\affiliation{Lunar \& Planetary Laboratory, University of Arizona, Tucson, AZ 85721 USA}
\affiliation{Habitability, Atmospheres, and Biosignatures Laboratory, University of Arizona, Tucson, AZ 85721, USA}
\affiliation{NASA Nexus for Exoplanet System Science Virtual Planetary Laboratory, University of Washington, Box 351580, Seattle, WA 98195, USA}

\author{Rhonda Morgan}
\affiliation{Jet Propulsion Laboratory, California Institute of Technology, 4800 Oak Grove Dr., Pasadena, CA, 91109, USA}

\author[0000-0002-8711-7206]{Dmitry Savransky}
\affiliation{Sibley School of Mechanical and Aerospace Engineering, Cornell University, Ithaca, NY, 14853, USA}
\affiliation{Carl Sagan Institute, Cornell University, Ithaca, NY, 14853, USA}

\author{Mario Damiano}
\affiliation{Jet Propulsion Laboratory, California Institute of Technology, 4800 Oak Grove Dr., Pasadena, CA, 91109, USA}

\author{Armen Tokadjian}
\affiliation{Jet Propulsion Laboratory, California Institute of Technology, 4800 Oak Grove Dr., Pasadena, CA, 91109, USA}

\begin{abstract}
Detection and characterization of Earth-like worlds is a primary science goal for the upcoming NASA flagship Habitable Worlds Observatory (HWO). The HWO mission will be the first of its kind to specifically search for Earth-like worlds around Sun-like stars and characterize their atmospheres to identify potential biosignatures. A mission with such ambitious science goals thus requires equally robust and physically realistic simulation tools. In this work, we extend the realism of the \texttt{EXOSIMS} framework by incorporating phase-dependent biosignature detectability. We use the spectral retrieval tool \texttt{rfast} to determine the signal-to-noise ratio (SNR) required to detect atmospheric O$_2$ and H$_2$O as a function of planetary phase and incorporate these requirements directly into \texttt{EXOSIMS} characterization integration-time calculations. Although the required molecular detection SNR decreases toward increasingly crescent phases, successful molecular detections are concentrated at intermediate phase angles, demonstrating the competing effects of molecular information content and direct-imaging observability. Relative to fixed characterization thresholds of SNR = 5 for H$_2$O and SNR = 8.5 for O$_2$, the phase-dependent treatment increases mean molecular detection yield by 76.8 percent for H$_2$O and 29.7 percent for O$_2$. Ultimately, we provide a framework that improves simulation fidelity, increases the realism of \texttt{EXOSIMS} as we expand to assess additional biosignatures in future work, and marries detection and characterization in yield modeling as we look to the future with HWO.
\end{abstract}

\section{Introduction} \label{sec:intro}
The Habitable Worlds Observatory (HWO) is an upcoming NASA flagship mission that will be a major step towards finding and characterizing Earth 2.0. The HWO mission will be the first of its kind to study Earth-like planets around Sun-like stars and assess their atmospheres for potential biosignatures. The Astro 2020 Decadal Survey emphasized not only the detection of Earth-like worlds, but also the characterization of their atmospheres through reflected-light spectroscopy to investigate their compositions and assess their potential habitability \citep{Astro2020}. Thus, the scientific return of HWO will also go beyond the number of Earth-like worlds we will discover by gaining information on the atmospheres of these worlds.

We have robust simulation tools that have proven useful for the detection goals of HWO; however, more work still needs to be done on the characterization goals of HWO and how we can marry the two. These exoplanet mission yield simulation tools provide a translation of these science objectives into observatory and mission requirements. The Exoplanet Open-Source Mission Simulator (\texttt{EXOSIMS}) \citep{Savransky2015, Delacroix2016, Morgan2017, savransky} and the Altruistic Yield Optimization \citep{stark2014maximizing} tool both model the expected yield of future direct imaging missions through testing various parameters including astrophysical assumptions, observatory structure, and observing requirements. The \texttt{EXOSIMS} model, as detailed in previous work \citep{Foote2026} performs end-to-end Monte Carlo mission simulations in which planetary systems are generated and observations are dynamically scheduled throughout a simulated mission. These frameworks have enabled increasingly sophisticated predictions of how many potentially habitable planets could be detected for a given mission architecture.

Spectral resolution, wavelength, SNR, and bandwidth are fundamental input parameters to mission-yield modeling. Previous studies have explored spectral characterization requirements, including the use of a continuum spectral SNR of 5 and resolving power of 70 as a characterization metric in \citet{stark2014maximizing}. For example, \citet{Morgan2022} evaluated both O$_2$ with a resolving power of 140 and a continuum spectral SNR of 8.5 as well as H$_2$O with a resolving power of 70 and an SNR of 5 \citep{morgan2019sdetfinal}. The water search requirement was motivated by atmospheric retrieval simulations of the 940\,nm H$_2$O absorption feature \citep{Feng2018}. More recently, \citet{stark2024optimized} explored wavelength optimization in yield calculations to determine optimal detection and characterization bandpasses, including for H$_2$O detection. These requirements propagate through the integration times and resulting yields. These metrics provide a computationally tractable connection between atmospheric science objectives and mission yield. However, simply counting a planet as spectrally detected and characterized according to a prescribed spectral-quality metric does not necessarily indicate whether a particular atmospheric molecule can be detected under the specific conditions of that observation.

Signal-to-noise ratio (SNR) is particularly important in connecting atmospheric detection requirements to mission yield because the integration time required for a characterization observation is a strong function of the target SNR (see equation 28 in \citep{nemati2020perf}. The SNR required for molecular detection also depends on the planet type and observing geometry. Determining a physically motivated SNR requirement is therefore important for assessing both whether a molecule can be detected and the mission resources required to obtain that detection.

An additional complication is that the information contained in a reflected-light spectrum is not necessarily constant throughout a planet's orbit. The observed spectrum depends on planetary phase, defined by the angle between the star, the planet, and the observer. Changes in illumination and viewing geometry alter both the apparent brightness of a planet and the atmospheric information contained in its reflected-light spectrum. In \citet{Foote2026}, we incorporated a realistic Earth phase function into \texttt{EXOSIMS} to replace the commonly adopted Lambertian approximation and found that phase-dependent brightness redistributes detections toward medium and larger phase angles. If the strength of a molecular absorption feature also changes with observing geometry, then the SNR required to detect that molecule may vary with phase. A single characterization SNR applied throughout the orbit, as \texttt{EXOSIMS} had previously assumed, therefore does not represent an equivalent molecular detection requirement at every phase. Here, we extend this phase-dependent treatment from planet detection to atmospheric characterization by making the molecular detection SNR phase dependent, connecting both effects within the same end-to-end mission simulation.

Previous studies have demonstrated the importance of planetary phase for reflected-light characterization, including phase-dependent spectral modeling and ocean-glint detection \citep{Ryan2022} and atmospheric retrieval studies examining the effects of orbital phase and phase constraints \citep{Nayak2017, Salvador2024}. Molecular detectability cannot be considered independently of the observing constraints of a direct imaging mission. A phase that provides a stronger molecular feature may simultaneously correspond to a fainter planet, a smaller projected star-planet separation, or otherwise less favorable observing conditions. Determining the observing geometries that ultimately provide successful molecular detections therefore requires the atmospheric detection requirement to be propagated through the entire simulation rather than considered independently so that we can connect this information to a specific mission such as HWO for Earth-like worlds.

In this work, we develop a generalized framework for incorporating phase-dependent molecular detection requirements directly into \texttt{EXOSIMS}. We use a computationally efficient planetary spectrum model to determine the spectral SNR required to detect H$_2$O and O$_2$ in reflected-light spectra of a modern-Earth analog as a function of planetary phase. These requirements are incorporated into \texttt{EXOSIMS} so that the molecular detection requirement at the phase of an observation directly informs the required characterization integration time. The achieved SNR can then be evaluated against the corresponding molecular requirement to determine whether a specific species would be detected. We apply this framework across ensembles of simulated HWO missions to ask how phase-dependent molecular detectability changes the phases at which atmospheric species can be successfully detected and the resulting mission-level molecular detection yield. In this way, we connect planet detection, atmospheric characterization, and species-specific molecular detectability within a common end-to-end mission simulation framework.

\section{Methods} \label{sec:methods}

Our approach connects atmospheric modeling with end-to-end mission simulations to determine how phase-dependent molecular detectability affects characterization yield. We first use \texttt{rfast} to determine the SNR required for molecular detection as a function of planetary phase and validate these requirements using atmospheric retrievals. We then implement the phase-dependent molecular detection requirements within \texttt{EXOSIMS}, where they are used to calculate characterization integration times and assess whether individual molecular species are successfully detected. Finally, we perform ensembles of simulated HWO missions to quantify how these phase-dependent requirements affect molecular characterization yield compared to fixed-SNR assumptions.

\subsection{\texttt{rfast} Molecular Detection Requirements}
We use the radiative transfer and atmospheric retrieval framework \texttt{rfast} \citep{RobinsonSalvador2023} to quantify the signal-to-noise ratio (SNR) required to detect individual atmospheric molecular species in reflected-light spectra of an Earth-like planet with a modern atmosphere. Instead of adopting a single characterization SNR requirement at all orbital phases, as previously assumed, we calculate a species-specific molecular detection requirement as a function of planetary phase angle. We consider H$_2$O and O$_2$ as initial cases because both are important targets for the atmospheric characterization of potentially habitable terrestrial planets and have prominent absorption features within the visible and near-infrared wavelength range considered for HWO.

\begin{table}[ht!]
\centering
\caption{\texttt{rfast} model parameters used for the H$_2$O and O$_2$ calculations.}
\label{tab:rfast_parameters}
\begin{tabular}{lll}
\hline
Category & Parameter & Value \\
\hline
Planet & Radius & $1.0\,R_{\oplus}$ \\
 & Mass & $1.0\,M_{\oplus}$ \\
 & Orbital distance & $1.0$ au \\
 & Surface flux albedo & 0.05 (grey) \\
 & Surface emissivity & $1.0$ \\
\hline
Star & Effective temperature & $5780$ K \\
 & Stellar radius & $1.0\,R_{\odot}$ \\
\hline
Atmosphere & Pressure levels & $50$ \\
 & Top pressure & $1$ Pa \\
 & Bottom pressure & $1.01\times10^{5}$ Pa \\
 & Temperature & $294$ K (isothermal) \\
 & N$_2$ VMR & $0.78$ \\
 & O$_2$ VMR & $0.21$ \\
 & H$_2$O VMR & $3\times10^{-3}$ \\
 & O$_3$ VMR & $7\times10^{-7}$ \\
 & CO$_2$ VMR & $4\times10^{-4}$ \\
 & CH$_4$ VMR & $2\times10^{-6}$ \\
 & Rayleigh scattering & Included \\
 & Collision-induced absorption & N$_2$, O$_2$ \\
\hline
Clouds & Cloud fraction & $0.50$ \\
 & Cloud-top pressure & $6\times10^{4}$ Pa \\
 & Cloud pressure thickness & $1\times10^{4}$ Pa \\
 & Reference optical depth & $10$ \\
 & Reference wavelength & $0.55\,\mu\mathrm{m}$ \\
\hline
Phase model & \texttt{rfast} source setting & \texttt{phas} \\
 & Disk-integration points & $100$ \\
\hline
H$_2$O & Wavelength range & $0.8$--$1.0\,\mu\mathrm{m}$ \\
 & Resolving power & $R=140$ \\
\hline
O$_2$ & Wavelength range & $0.6$--$0.8\,\mu\mathrm{m}$ \\
 & Resolving power & $R=140$ \\
\hline
\label{tbl:rfast_params}
\end{tabular}
\end{table}

Table~\ref{tbl:rfast_params} shows the \texttt{rfast} model parameters used for the O$_2$ and H$_2$O calculations. For these calculations, we adopt the standard \texttt{rfast} setup for Modern Earth, with a radius and mass of 1 $R_\oplus$ and 1 $M_\oplus$, respectively, orbiting at 1\,AU from a Sun-like star with an effective temperature of 5780 K. The atmosphere was represented using 50 pressure levels spanning 1 Pa to $1.01\times10^5$ Pa and included molecular absorption, Rayleigh scattering, and clouds. The adopted model included a cloud fraction of 0.5. The H$_2$O volume mixing ratio was $3\times10^{-3}$, while the O$_2$ volume mixing ratio was 0.21.

We generated reflected-light spectra at a spectral resolving power of $R=140$. Water vapor was evaluated over $0.8$--$1.0~\mu$m, encompassing the strong H$_2$O absorption feature near $0.94~\mu$m, while O$_2$ was evaluated over $0.6$--$0.8~\mu$m, encompassing the O$_2$ A-band near $0.76~\mu$m. These wavelength intervals were chosen to correspond to the characterization bands subsequently represented in the \texttt{EXOSIMS} simulations. For each species, spectra were generated at phase angles of

$$ \alpha = [0^\circ,15^\circ,30^\circ,60^\circ,90^\circ,110^\circ,130^\circ,150^\circ,175^\circ]. $$

\noindent Here, $\alpha$ is the star-planet-observer angle, such that $\alpha=0^\circ$ corresponds to full phase, $90^\circ$ to quadrature, and increasing phase angles beyond quadrature correspond to increasingly crescent phases.

At each phase angle, we generate paired spectra in which the molecular species of interest was either present at its nominal modern-Earth abundance or effectively removed from the atmosphere. All other parameters were held fixed between each pair of calculations. The resulting molecule-present and molecule-removed spectra therefore isolate the spectral contribution of the species of interest at each planetary phase. 

For each phase angle, the molecular signature was quantified from the wavelength-dependent difference between the molecule-present and molecule-removed spectra. Across the bandpass, the ``signal'' then varies with wavelength according to the combination of the planetary reflectance spectrum, the molecular difference signature, and square of the wavelength, as shown in \citet{robinson2016coronagraph}, their Equation~14. Similarly, the ``noise'' (in terms of photons) was assumed to vary with wavelength to the fourth power, which is the most conservative assumption and that is appropriate for a scene dominated by exozodiacal background (\citep{robinson2016coronagraph}, their Equation 18). Given a desired confidence for the molecular detection, the wavelength-scaled SNR is integrated over the spectral dimension to yield a single, requisite spectral SNR, referred to in the rest of this paper as SNR$_{\rm req}$ and usually specified at the shortest wavelength in the bandpass. In other words, the resulting requirement represents the spectral SNR normalization necessary for detection of the molecular feature across the full wavelength interval, rather than the significance of a single spectral element. We adopt a $5\sigma$ molecular detection criterion throughout this work. Repeating this calculation over the complete phase grid produces separate phase-dependent SNR$_{\rm req}(\alpha)$ relations for H$_2$O and O$_2$. For each molecular species, the calculated phase angles and corresponding SNR$_{\rm req}$ values were stored in their own lookup table for use by \texttt{EXOSIMS}. These tables provide the connection between the atmospheric spectral calculations and the mission simulation. 

\begin{figure*}[ht!]
\centering
\includegraphics[width=\textwidth]{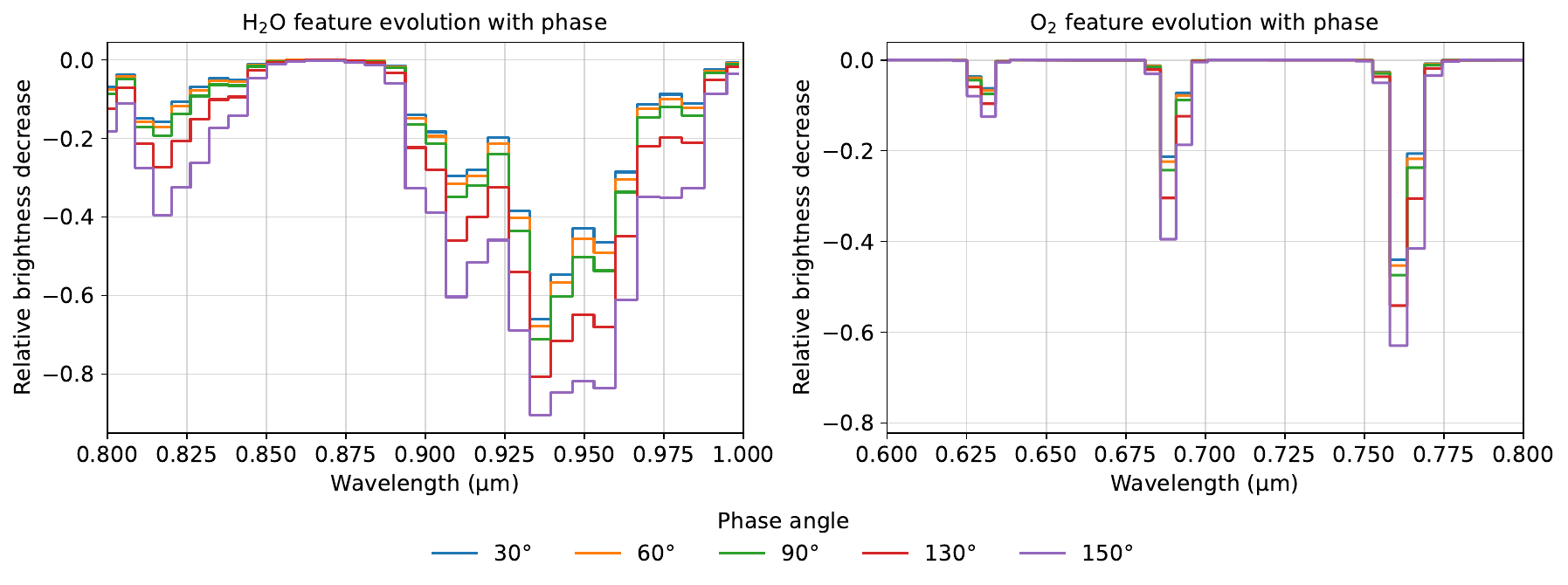}
\caption{Phase dependence of the H$_2$O (left) and O$_2$ (right) molecular absorption features used to determine the molecular detection requirements. Curves show the relative spectral difference between the molecule-present and molecule-removed \texttt{rfast} spectra at representative planetary phase angles. H$_2$O is shown over $0.8$--$1.0~\mu$m and O$_2$ over $0.6$--$0.8~\mu$m, both at $R=140$. The magnitude of the molecular absorption features generally increases toward larger phase angles, demonstrating that the relative molecular spectral signal varies with observing geometry.}
\label{fig:feature_phase}
\end{figure*}

The molecular absorption signals of both H$_2$O and O$_2$ vary with planetary phase angle (Figure~\ref{fig:feature_phase}). For H$_2$O, the absorption features across the 0.8--1.0~$\upmu$m wavelength interval become increasingly pronounced toward larger phase angles, with particularly strong changes in the feature near 0.94~$\upmu$m. A similar phase dependence is present for O$_2$ across the $0.6$--0.8~$\upmu$m interval, including the O$_2$ A-band near 0.76~$\upmu$m. These results demonstrate that the relative molecular spectral signal is not constant throughout the planet's orbit.

\begin{figure}[ht!]
\centering
\includegraphics[width=\columnwidth]{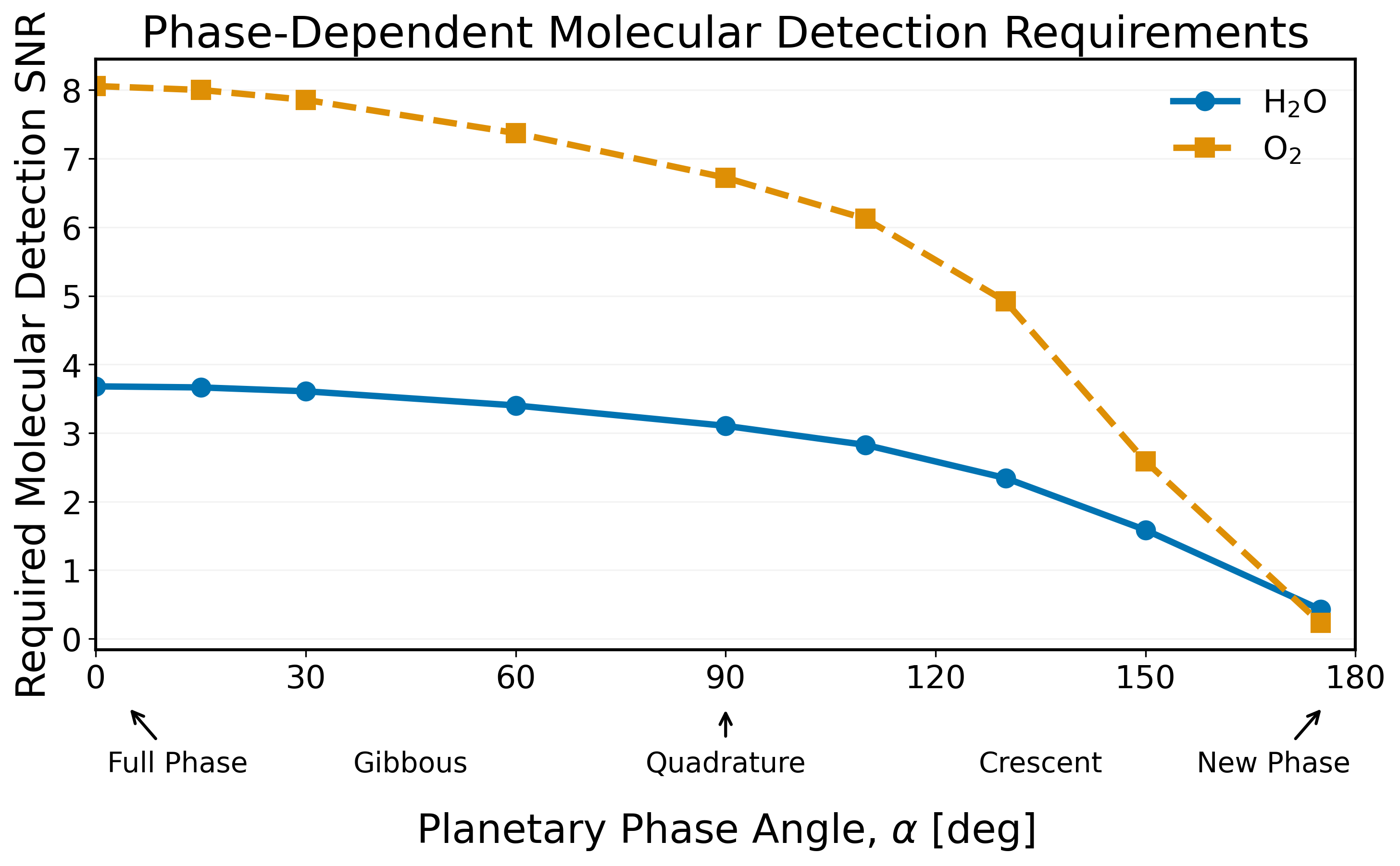}
\caption{Required molecular detection SNR, SNR$_{\rm req}$, as a function of planetary phase angle for H$_2$O and O$_2$. Values are derived from paired molecule-present and molecule-removed \texttt{rfast} spectra using a $5\sigma$ molecular detection criterion. The required SNR decreases with increasing phase angle for both species. Lines connect the discrete \texttt{rfast} calculations used to construct the phase-dependent lookup tables incorporated into \texttt{EXOSIMS}.}
\label{fig:snr_phase}
\end{figure}

The phase dependence of the molecular spectral features produces a corresponding phase dependence in the SNR required for molecular detection (Figure~\ref{fig:snr_phase}). For H$2$O, SNR${\rm req}$ decreases from 3.68 at full phase ($\alpha=0^\circ$) to 3.10 at quadrature and 1.59 at $\alpha=150^\circ$. For O$_2$, the corresponding requirements decrease from 8.06 at full phase to 6.72 at quadrature and 2.59 at $\alpha=150^\circ$. At the largest sampled phase angle of $175^\circ$, the required SNR reaches 0.43 for H$_2$O and 0.23 for O$_2$. Across the sampled phase grid, the molecular detection requirement therefore decreases monotonically with increasing phase angle for both species.

\subsection{Validation of the Molecular Detection Requirement}

To test whether the semi-analytic molecular detection requirement provides a reasonable approximation to full atmospheric retrieval performance, we performed \texttt{rfast} retrieval calculations for both H$_2$O and O$_2$ at quadrature ($\alpha=90^\circ$) over a range of input spectral SNR values. For H$_2$O, retrievals were evaluated at SNR values from 1 to 5. The H$2$O abundance was poorly constrained at the lowest SNR values, with the constraint developing at SNR values of approximately 3--5. This behavior is consistent with the semi-analytic $5\sigma$ molecular detection requirement of SNR${\rm req}=3.10$ at quadrature. For O$2$, retrievals were evaluated at SNR values of 2, 4, 6, 8, and 10, using 150,000 sampling steps for the final calculations. Oxygen was not constrained at the lowest SNR values, became detectable between SNR $\sim6$ and 8, and was clearly constrained at the highest SNR values. This behavior is similarly consistent with the semi-analytic $5\sigma$ molecular detection requirement of SNR${\rm req}=6.72$ at quadrature. Together, the retrieval results support the use of the precomputed SNR$_{\rm req}(\alpha)$ relations as computationally efficient proxies for molecular detectability within the mission simulation. 

\subsection{Phase-Dependent Molecular SNR Framework in \texttt{EXOSIMS}}

To incorporate \texttt{rfast}-derived molecular detection requirements into \texttt{EXOSIMS}, we develop a modular phase-dependent molecular SNR framework consisting of three components: \texttt{SNRLookup}, \texttt{PhaseDependentSNR}, and a modified scheduler \texttt{coroOnlyScheduler}. Please see \citet{Foote2026} for a detailed description of \texttt{EXOSIMS}. \texttt{SNRLookup} acts as a wrapper that handles loading and interpolation of the molecular detection lookup tables, \texttt{PhaseDependentSNR} subclasses the existing \texttt{Nemati} \texttt{OpticalSystem} and provides the interface between these lookup tables and the \texttt{EXOSIMS} integration time calculation, and the modified scheduler \texttt{coroOnlyScheduler} supplies the instantaneous planetary phase angle and records the resulting molecular detection information. This structure preserves the existing \texttt{EXOSIMS} integration-time calculation while allowing its characterization SNR requirement to vary with molecular species and planetary phase.

When \texttt{EXOSIMS} considers a characterization observation, the scheduler determines the instantaneous planetary phase angle from the simulated orbital geometry. This phase angle is then passed to \texttt{PhaseDependentSNR}, which identifies the molecular species associated with the applicable characterization observing mode. \texttt{SNRLookup} converts that phase angle from radians to degrees and linearly interpolates between the tabulated \texttt{rfast} calculations to obtain SNR$_{\rm req}(\alpha)$. 

For characterization modes to which the molecular lookup applies, the interpolated SNR$_{\rm req}(\alpha)$ temporarily replaces the fixed SNR stored in the observing mode. The modified observing mode is then passed to the existing \texttt{Nemati.calc\_intTime()} calculation, leaving the underlying \texttt{EXOSIMS} count-rate and noise model unchanged. Thus, the atmospheric calculation modifies the required characterization SNR that is handed to the integration time calculation, rather than replacing the \texttt{EXOSIMS} integration-time model itself. Since the required integration time strongly depends on the target SNR, this framework allows phase-dependent molecular information requirements to propagate directly into the amount of mission time allocated to characterization. 

Molecular requirements are applied only to the characterization observing mode associated with the relevant spectral feature. H$_2$O and O$_2$ are therefore mapped to their respective $R=140$ characterization configurations, while detection observations and related observing modes retain their standard \texttt{EXOSIMS} SNR requirements. If molecular lookup functionality is disabled or no configured molecular species applies to a given mode, \texttt{PhaseDependentSNR} returns the original observing-mode SNR and \texttt{EXOSIMS} proceeds through its standard characterization workflow.

Although H$_2$O and O$_2$ are evaluated individually in the simulations presented here, the framework is designed to support multiple molecular species and observing modes. Separate lookup tables may be supplied for each species, and the species applicable to each characterization mode are specified in the mission configuration. When multiple species are associated with a single mode, the framework can select the maximum, minimum, or first molecular SNR requirement according to a user-specified strategy. The present calculations use a single species per simulation, so this choice does not affect our results reported here. 

Following each characterization observation, the achieved characterization SNR is compared with the molecular requirement evaluated at the same planetary phase angle. For a species, we classify the molecule as detected when SNR$_{\rm achieved} \geq SNR_{\rm req}(\alpha)$. The molecular species evaluated, required SNR, planetary phase angle, and resulting detection flag are then recorded in the Design Reference Mission (DRM). \texttt{EXOSIMS} does not perform a new atmospheric retrieval during each simulated observation. Instead, the \texttt{rfast} calculations provide a precomputed mapping between the phase angle and the SNR corresponding to a $5\sigma$ molecular detection, which is then evaluated within the mission simulation.

We evaluated the phase-dependent molecular characterization framework using ensembles of \texttt{EXOSIMS} mission simulations representative of an HWO direct-imaging survey. The baseline configuration assumed a 6 m telescope and a five-year mission lifetime. The adopted coronagraph model had an inner working angle of 0.054 arcsec, an outer working angle of 0.422 arcsec, and a raw contrast of $10^{-10}$. Planetary occurrence rates followed \citet{dulz2019exoplanet}, using a geometric albedo of 0.2. Other astrophysical parameters followed the Exo-Earth Candidate description used by HabEx, LUVOIR, and HWO, as described in the ExEP Standards Team Final Report \citep{morgan2019sdetfinal}. Observations were scheduled using \texttt{coroOnlyScheduler}. The simulations did not use the \texttt{lucky\_planets} approximation, allowing planetary orbital geometry and phase to evolve self-consistently during the mission.

Planet detection was performed at 550 nm with a photometric detection SNR requirement of 7, independently of the molecular lookup framework. Atmospheric characterization used an $R=140$ spectroscopic mode. For H$_2$O, the characterization mode was centered at 910\,nm, while the O$_2$ configuration covered the 600--800\,nm spectral region represented by the \texttt{rfast} calculations. Consequently, the molecular framework modifies characterization requirements without changing the criteria used by \texttt{EXOSIMS} to initially detect a planet.

Separate ensembles of 1000 simulated missions were performed for H$_2$O and O$_2$ using their respective phase-dependent \texttt{rfast} lookup tables. Each realization represents an independent simulated five-year mission. For every attempted applicable characterization observation, the DRM records the phase angle, achieved characterization SNR, required molecular SNR, and molecular detection outcome. These quantities allow molecular detections to be evaluated both as a function of observing geometry and at the mission-yield level.

To isolate the mission-level consequences of replacing a fixed molecular SNR requirement with a phase-dependent requirement, we additionally performed fixed-SNR control simulations using the same molecular lookup architecture. The phase-dependent lookup tables were replaced with constant-valued tables of SNR = 5 for H$_2$O and SNR = 8.5 for O$_2$, corresponding to the molecular SNR thresholds adopted by Morgan et al. (2022). These controls retain the observing configurations used in our phase-dependent simulations, including $R=140$ characterization modes; they are therefore not intended as exact reproductions of the Morgan et al. simulations. All other simulation and molecular-detection bookkeeping remained unchanged. Differences between the control and phase-dependent ensembles therefore quantify the mission-level consequences of the adopted molecular SNR treatment, including its effects on characterization integration time, scheduling, and molecular detection yield.

We quantify atmospheric characterization performance using both individual characterization outcomes and mission-level molecular yield. A molecular characterization attempt is defined as an observation for which the applicable molecular characterization mode is considered by the scheduler. A successful molecular characterization requires both that the achieved characterization SNR satisfies the corresponding species-specific molecular threshold and that the observation meets the characterization success criterion. Thus, satisfying the molecular SNR threshold alone does not necessarily contribute to the reported molecular detection yield. We examine the number of successful molecular detections as a function of phase as well as the molecular detection fraction,
\begin{equation}
f_{\rm det}(\alpha) =
\frac{N_{\rm mol,det}(\alpha)}
     {N_{\rm mol,attempt}(\alpha)} \ ,
\end{equation}
which accounts for the nonuniform number of characterization opportunities occurring at different planetary phases.

At the mission level, we define the molecular detection yield as the number of successful detections of a specified molecular species during a single simulated mission. We report the resulting yield distributions across the ensemble of mission realizations and compare the phase-dependent and fixed-SNR cases. This metric extends the conventional use of planet detection or characterization yield by explicitly requiring that the simulated characterization observation achieve the information quality associated with the detection of a particular atmospheric molecule.

\section{Results} \label{sec:results}

\begin{figure}[!htbp]
\centering
\includegraphics[width=\columnwidth]{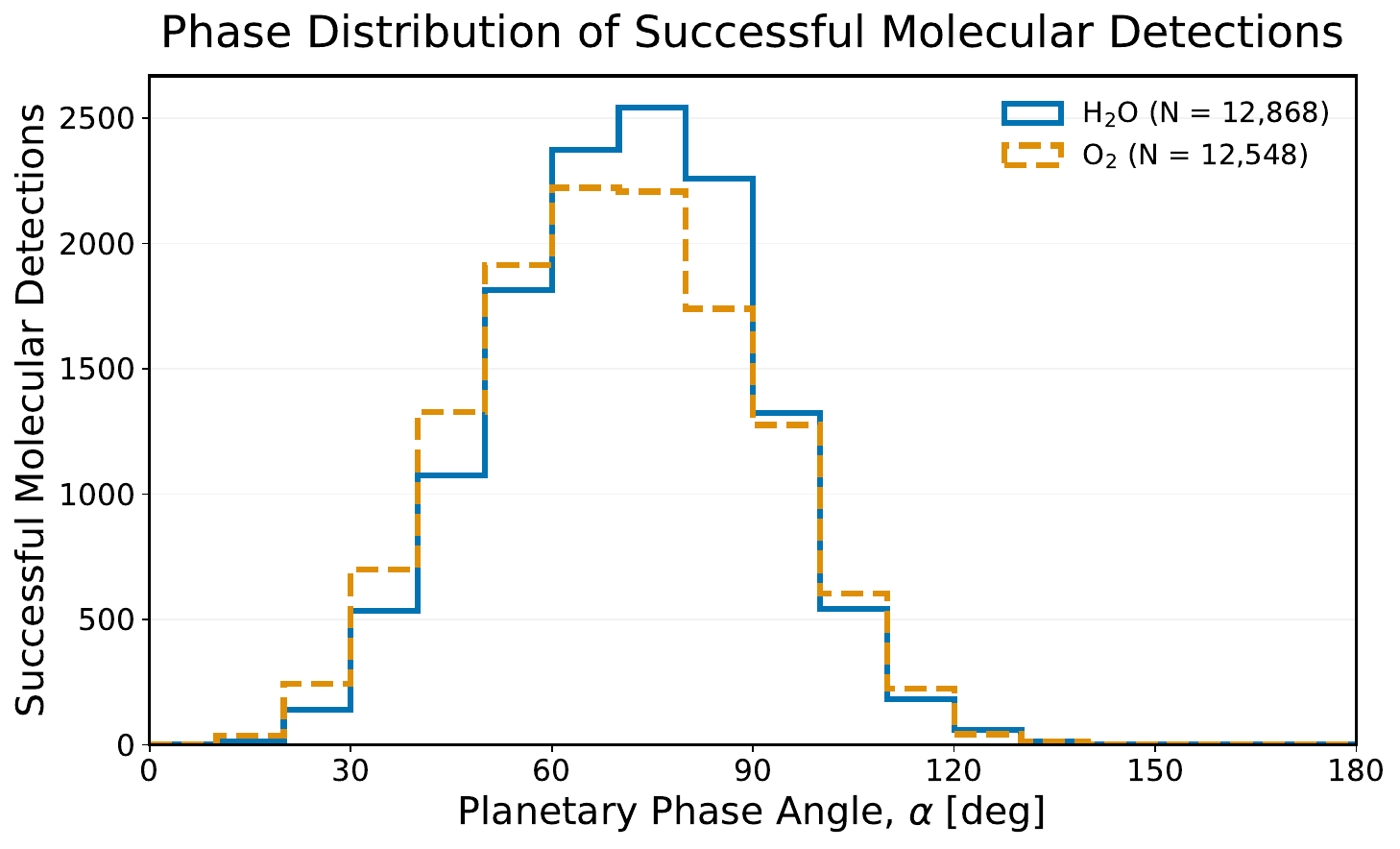}
\caption{Phase distribution of successful H$_2$O and O$_2$ molecular detections across 1000 simulated five-year missions. Successful detections are shown in $10^\circ$ bins of planetary phase angle. The ensembles contain 12,868 successful H$_2$O detections and 12,548 successful O$_2$ detections. For both species, successful detections are concentrated at intermediate phase angles rather than at the largest phase angles, where the \texttt{rfast}-derived molecular detection requirements are lowest.}
\label{fig:detection_counts}
\end{figure}

Despite the decreasing molecular detection requirements toward larger phase angles, successful molecular detections in the \texttt{EXOSIMS} mission simulations are concentrated at intermediate phases (Figure~\ref{fig:detection_counts}). Across the 1000 simulated missions, 12,868 successful H$_2$O detections and 12,548 successful O$_2$ detections are recorded. For both species, the number of successful detections increases from near-full phases, reaches its largest values at intermediate phase angles, and then declines toward increasingly crescent phases. Thus, the phase angles associated with the lowest \texttt{rfast}-derived SNR$_{\rm req}$ do not correspond to the phase angles at which the largest number of molecular detections occur.

We use ensembles of 1000 simulated missions to reduce Monte Carlo sampling noise and better characterize the resulting phase and yield distributions. Because planets naturally evolve through their orbits rather than being forced to a uniform phase distribution, the number of characterization attempts differs among phase bins. We therefore report molecular detection fractions with 68.3\% Wilson binomial confidence intervals, which account for the finite and unequal number of attempts in each bin. The large ensemble improves the statistical precision of these distributions but does not remove uncertainties associated with the underlying atmospheric and mission assumptions.

\begin{figure}[!htbp]
\centering
\includegraphics[width=\columnwidth]{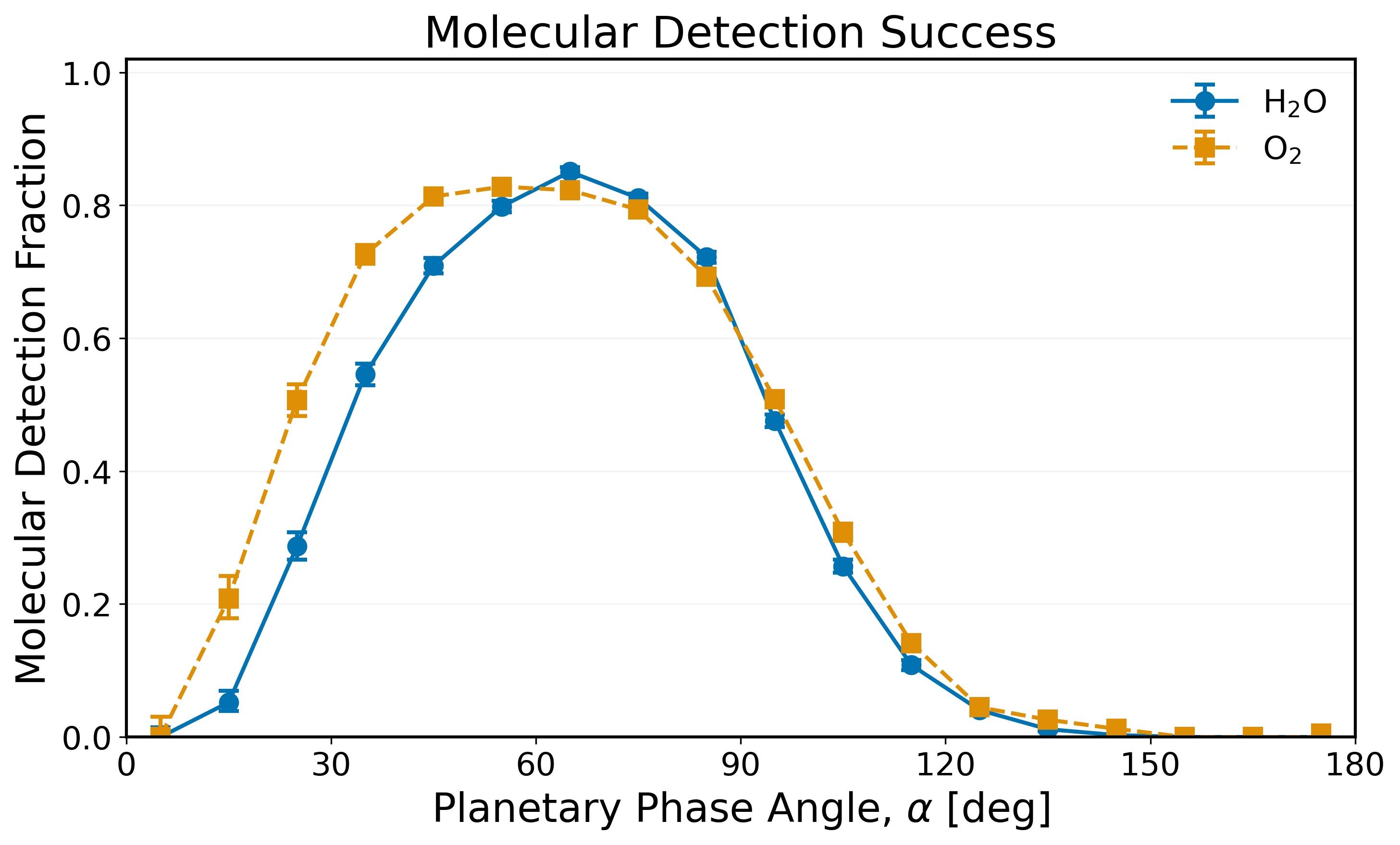}
\caption{Molecular detection fraction as a function of planetary phase angle for H$_2$O and O$_2$ across 1000 simulated missions. The detection fraction is defined as the number of successful molecular detections divided by the number of molecular characterization attempts within each $10^\circ$ phase bin. Error bars show 68.3\% binomial confidence intervals. Detection success for both species reaches a maximum at intermediate phase angles before declining through quadrature and toward increasingly crescent phases.}
\label{fig:detection_fraction}
\end{figure}

Normalizing the successful detections by the number of molecular characterization attempts reveals a similar phase dependence for H$_2$O and O$_2$ (Figure~\ref{fig:detection_fraction}). For both species, the molecular detection fraction rises from near-full phases and reaches a maximum at intermediate phase angles. The detection fraction then declines through quadrature and toward increasingly crescent phases. This behavior is notable because SNR$_{\rm req}$ continues to decrease toward larger phase angles. The molecular detection probability therefore does not simply track the intrinsic \texttt{rfast}-derived molecular detection requirement, indicating that the observing conditions encountered by \texttt{EXOSIMS} also play an important role in determining whether a molecular characterization is successful.

\begin{figure*}[ht!]
\centering
\includegraphics[width=\textwidth]{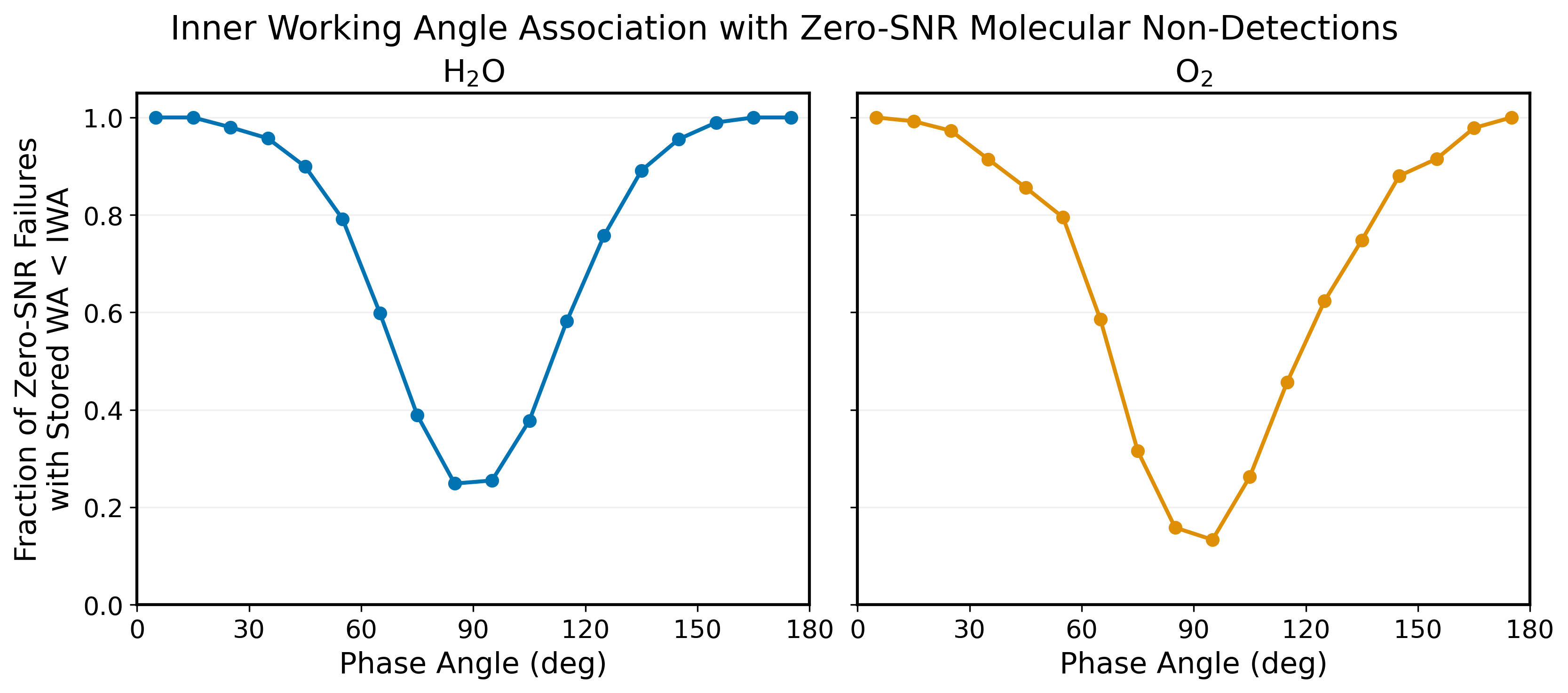}
\caption{Association between the coronagraph inner working angle (IWA) and zero-SNR molecular non-detections as a function of planetary phase angle for H$_2$O (left) and O$_2$ (right). For molecular non-detections with an achieved characterization SNR of zero, the plotted fraction indicates the proportion for which the stored planet working angle (WA) falls interior to the coronagraph IWA ($WA < IWA$), where the planet is inaccessible to the coronagraph. The increasing association with the IWA toward larger phase angles indicates the growing importance of coronagraph accessibility in limiting molecular characterization at increasingly crescent geometries. This effect is more pronounced for H$_2$O, which is characterized at longer wavelengths, resulting in a larger IWA due to its dependence on $\lambda/D$.}
\label{fig:iwa_failures}
\end{figure*}

To further investigate the decline in molecular detection success toward larger phase angles, we examined molecular non-detections for which the achieved characterization SNR was zero. Figure~\ref{fig:iwa_failures} shows the fraction of these zero-SNR non-detections for which the stored planet working angle falls interior to the coronagraph IWA. For both H$_2$O and O$_2$, the association between zero-SNR non-detections and the IWA increases toward larger phase angles. This indicates that coronagraph accessibility becomes increasingly important at these observing geometries, even as the intrinsic \texttt{rfast}-derived molecular SNR requirement continues to decrease.

\begin{figure}[ht!]
\centering
\includegraphics[width=\columnwidth]{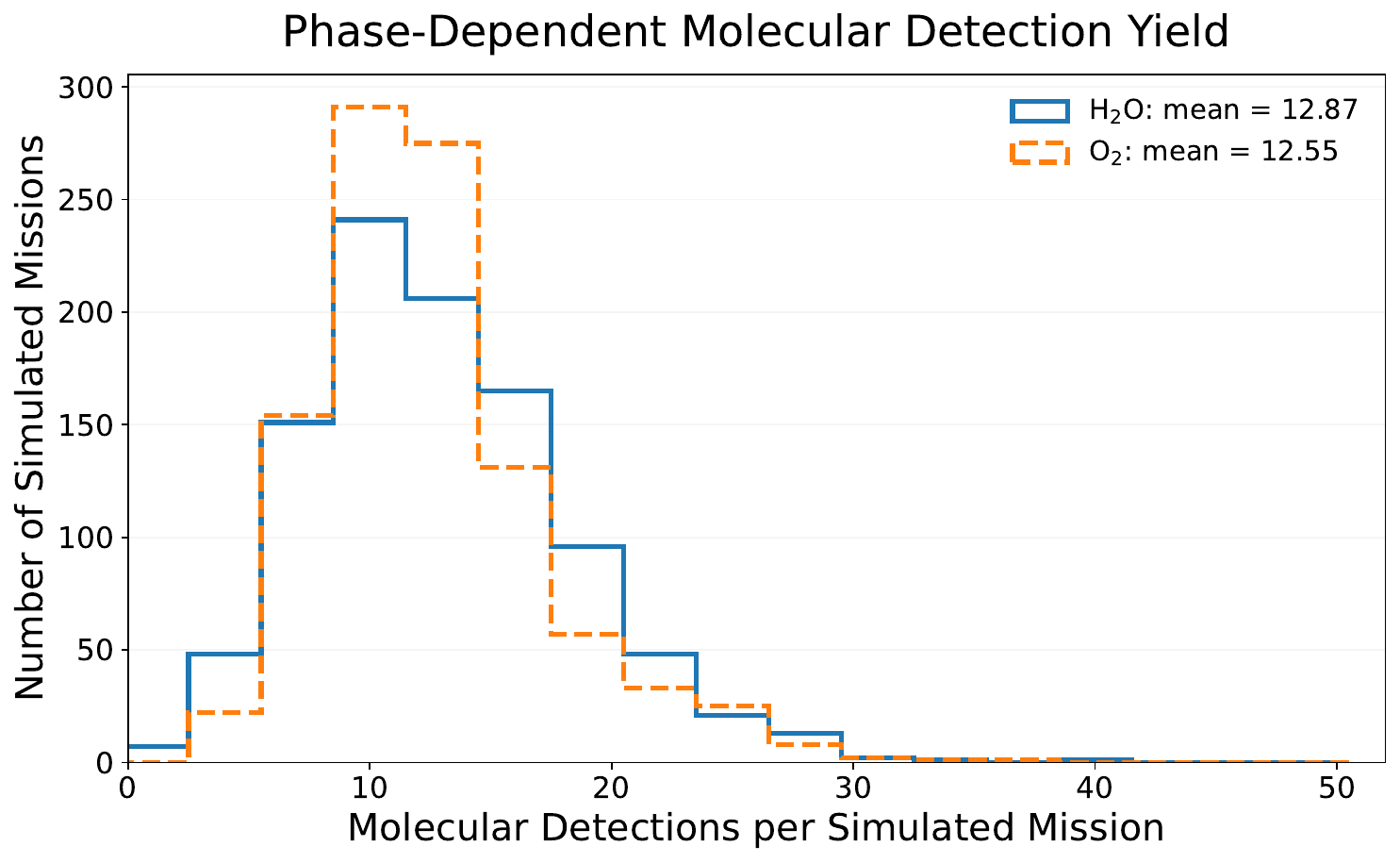}
\caption{Distribution of molecular detection yield across 1000 simulated five-year missions using the phase-dependent molecular detection requirements. Molecular detection yield is defined as the number of successful detections of the specified species within an individual simulated mission. The mean yields are 12.9 H$_2$O detections and 12.6 O$_2$ detections per mission.}
\label{fig:molecular_yield}
\end{figure}

The phase-dependent molecular characterization framework produces distinct mission-level molecular detection yield distributions for H$_2$O and O$_2$ (Figure~\ref{fig:molecular_yield}). Across the 1000 simulated missions, the mean H$_2$O molecular detection yield is 12.9 detections per mission, while the mean O$_2$ molecular detection yield is 12.6 detections per mission. The mission-to-mission distributions illustrate the range of molecular detection outcomes produced across independent realizations of the same survey architecture.

\begin{figure*}
    \centering
    \includegraphics[width=\textwidth]
    {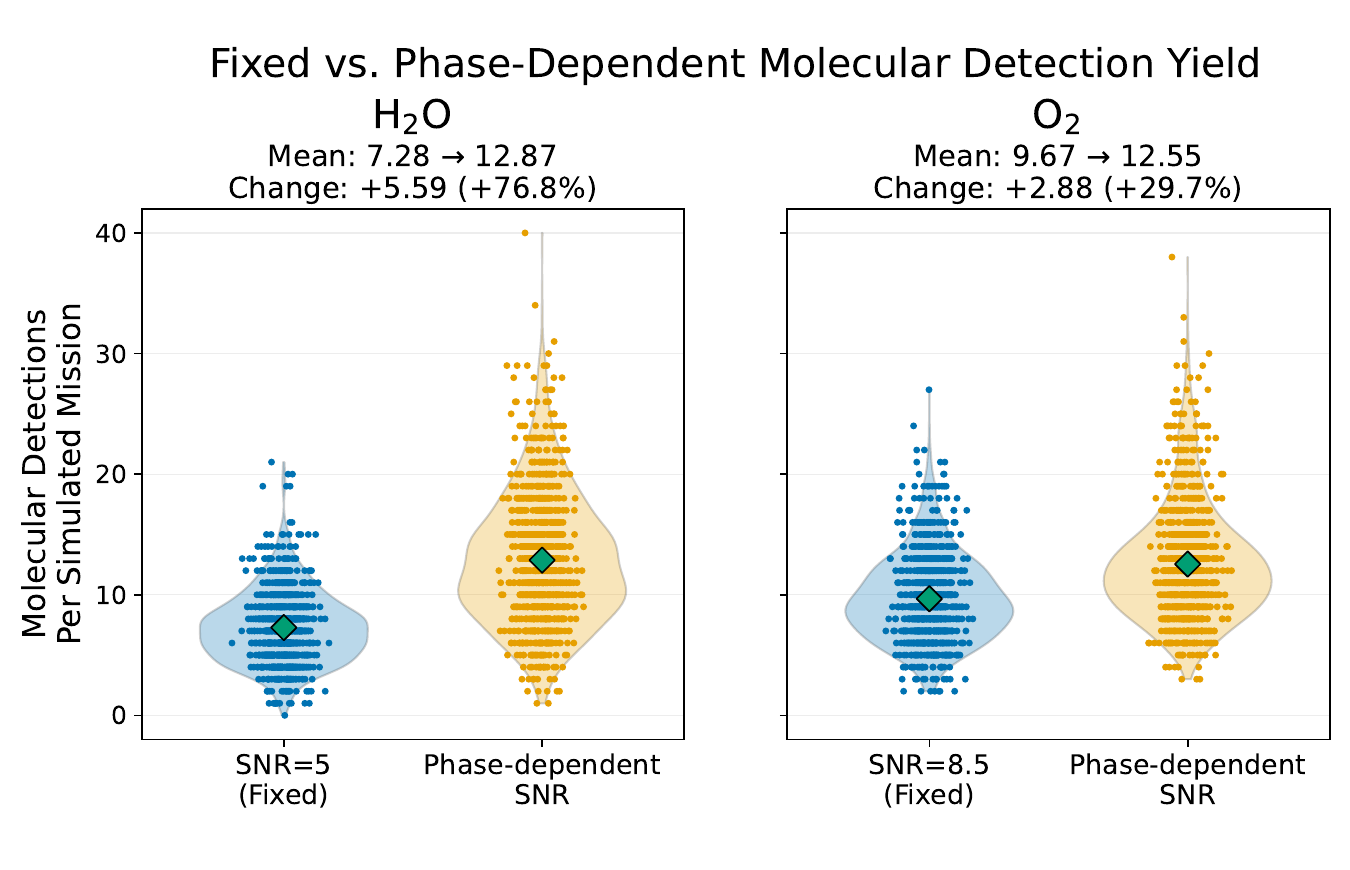}
   \caption{Distribution of molecular detection yields for H$_2$O and O$_2$ under phase-dependent and fixed-SNR treatments. The phase-dependent treatment uses the molecular detection thresholds derived from the \texttt{rfast} phase-dependent SNR calculations, while the fixed-SNR control adopts constant molecular SNR thresholds of 5 for H$_2$O and 8.5 for O$_2$, following the values used by \citet{Morgan2022}, within the same \texttt{EXOSIMS} observing configuration. Violin widths indicate the relative density of simulated mission yields, with the internal markers showing the central tendency and spread of each distribution. Relative to the fixed-SNR treatment, the phase-dependent treatment increases the mean molecular detection yield by 76.8 percent for H$_2$O and 29.7 percent for O$_2$.}
    \label{fig:fixed_vs_phase}
\end{figure*}

The phase-dependent treatment increases molecular detection yield relative to the fixed-SNR treatment for both species (Figure~\ref{fig:fixed_vs_phase}). For H$_2$O, the mean molecular detection yield increases from 7.3 detections per mission in the fixed-SNR ensemble to 12.9 detections per mission with the phase-dependent treatment, an increase of 76.8 percent. For O$_2$, the mean yield increases from 9.7 to 12.6 detections per mission, corresponding to a 29.7 percent increase. Because the underlying mission configuration is held fixed between the two treatments, these differences demonstrate that the adopted molecular detection requirement can affect predicted atmospheric characterization yield.

\section{Discussion} \label{sec:discussion}

The \texttt{rfast} calculations demonstrate that the atmospheric information available in a reflected-light spectrum varies with planetary phase. As the planet moves toward larger phase angles, the relative H$_2$O and O$_2$ absorption features become stronger in the adopted models, and the spectral SNR required for a $5\sigma$ molecular detection decreases. A fixed characterization SNR therefore does not represent an equivalent molecular detection requirement at all observing geometries. By incorporating these phase-dependent requirements directly into \texttt{EXOSIMS}, this work connects the changing information content of the atmosphere to the observing time and molecular detection outcomes predicted over an entire mission.

Despite the progressively lower molecular SNR requirements toward larger phase angles, \texttt{EXOSIMS} does not show a corresponding monotonic increase in molecular detection success at these phases. Instead, successful H$_2$O and O$_2$ detections are concentrated at intermediate phase angles, and the normalized molecular detection fraction reaches a maximum near $60$--$70^\circ$. This distinction is crucial because the raw number of molecular detections depends on where the mission actually obtains characterization opportunities. Normalizing by the number of characterization attempts shows that the intermediate phase preference persists even after accounting for this nonuniform sampling of orbital phase. 

The intermediate-phase maximum reflects a competition between atmospheric information content and direct-imaging observability. At larger phase angles, the molecular features become stronger relative to the molecule-removed spectrum, reducing the spectral SNR required to identify the species. At the same time, however, the illuminated fraction of the planet decreases, reducing the available reflected-light signal, while the projected angular separation between the planet and host star evolves with orbital geometry. The phase that minimizes the intrinsic molecular detection requirement therefore need not minimize the observing time or maximize the probability of a successful characterization. Instead, the molecular detection yield emerges from the combination of the atmospheric requirement, planetary brightness, angular separation, coronagraph accessibility, and the scheduling decisions made throughout the mission.

These results extend the role of planetary phase in direct-imaging yield calculations beyond planet detection alone. In our previous work, replacing a Lambertian phase approximation with a realistic Earth phase function altered the apparent planetary brightness and redistributed \texttt{EXOSIMS} detections across orbital phase \citep{Foote2026}. The present work demonstrates that phase also affects atmospheric characterization after a planet has been detected by changing the spectral information required to identify a molecular species. Planetary phase therefore enters the mission problem in two distinct but coupled ways: it affects whether and when a planet can be observed, and it affects the atmospheric information obtainable from the resulting spectrum. Treating both effects within the mission simulation provides a more physically motivated connection between exo-Earth detection and atmospheric characterization.

The characterization failure analysis provides additional evidence for this interpretation. Molecular non-detections with zero achieved characterization SNR become increasingly associated with working angles interior to the coronagraph IWA toward larger phase angles (Figure 5). Thus, the decline in molecular detection fraction at increasingly crescent phases cannot be understood from the molecular SNR requirement alone. Although the relative molecular spectral signal becomes more favorable at these phases, the changing projected star--planet separation increasingly limits the geometries accessible to the coronagraph. The intermediate-phase maximum therefore emerges in part from the competing effects of improving molecular information content and increasingly restrictive direct-imaging observability.

The fixed-SNR control simulations demonstrate that this distinction has measurable consequences for predicted atmospheric science yield. When the phase-dependent molecular requirements are replaced by constant SNR thresholds while retaining the same \texttt{EXOSIMS} observing configuration, the resulting molecular detection yields are lower for both H$_2$O and O$_2$, although the magnitude of the difference varies between species. Because characterization integration time depends strongly on the adopted target SNR, changing the molecular requirement affects not only whether an individual observation satisfies the molecular detection criterion, but also the amount of mission time allocated to that characterization and the observing opportunities available later in the survey. The difference between the fixed and phase-dependent ensembles therefore reflects the propagation of the atmospheric information requirement through the end-to-end mission simulation, rather than a post-processing reclassification of otherwise identical observations. Even though relatively few successful molecular detections occur at high phase angles, the reduced SNR demand for these detections can free up mission time compared with requiring a fixed SNR of 5 for H$_2$O or 8.5 for O$_2$ at crescent phases. This reduction in the required characterization SNR can decrease the time cost of individual characterization attempts, allowing that mission time to be used for additional observing opportunities.

The magnitude of this effect differs between H$_2$O and O$_2$. For H$_2$O, the phase-dependent treatment produces more than twice as many characterization attempts on average as the fixed-SNR treatment, while increasing the mean molecular detection yield from 7.3 to 12.9 detections per mission. For O$_2$, the number of characterization attempts increases more modestly, while the total characterization time remains similar between the fixed and phase-dependent treatments. The resulting increase in O$_2$ molecular detection yield is therefore smaller, from 9.7 to 12.6 detections per mission. These differences illustrate that the mission-level impact of a phase-dependent molecular requirement depends on how the species-specific SNR requirement interacts with the observing configuration and mission schedule.

This framework also introduces a more science-specific definition of mission yield. Traditional direct-imaging yield calculations quantify quantities such as the number of planets detected or the number receiving successful characterization observations. These metrics remain essential, but they do not directly specify whether a particular atmospheric species would be detectable in the resulting spectrum. Here, the atmospheric requirement derived from \texttt{rfast} determines the characterization SNR target that propagates through \texttt{EXOSIMS}, and the achieved observation is subsequently evaluated against that same species-specific requirement. This enables a molecular detection yield to be calculated across complete simulated missions, providing a direct connection between planet detection, atmospheric characterization, and the species-specific science objectives of HWO.

An important consequence is that the observing geometry most favorable for atmospheric characterization cannot be identified from molecular feature strength alone. A phase angle with a comparatively low molecular SNR requirement may still be inefficient or inaccessible once planet brightness, projected separation, coronagraph performance, and integration time are considered. Conversely, the optimal characterization geometry may occur where the molecular feature itself is not maximized. This coupling suggests that architecture trades involving inner working angle, wavelength coverage, throughput, and characterization strategy should be evaluated using atmospheric-science metrics propagated through the mission simulation rather than using molecular detectability or planet accessibility in isolation.

The broader utility of this framework is that HWO architecture studies can be evaluated increasingly close to the atmospheric science objectives that motivate the mission. Future work will extend this approach beyond a modern-Earth atmosphere to atmospheric states representative of different stages of Earth's evolution and to additional molecular species and biosignatures of interest. The modular implementation also enables future studies of how parameters such as wavelength coverage, spectral resolution, throughput, and inner working angle affect species-specific atmospheric science return. Ultimately, connecting planet detection, characterization requirements, and molecular detection yield provides a path toward evaluating HWO architectures not only by the number of worlds they discover, but by the atmospheric science those observations can deliver.

\section{Conclusions} \label{sec:conclusions}

We developed a generalized framework that incorporates phase-dependent, species-specific molecular detection requirements derived from \texttt{rfast} directly into \texttt{EXOSIMS} mission simulations. This approach connects the orbital geometry that determines whether an exo-Earth can be observed with the atmospheric information that can be obtained from its characterization spectrum.

\begin{itemize}

\item The relative H$_2$O and O$_2$ molecular absorption signals vary with planetary phase, producing phase-dependent SNR requirements for a $5\sigma$ molecular detection.

\item Although the required molecular detection SNR decreases toward larger phase angles, molecular detection success does not increase monotonically. Instead, the simulations show a sweet spot at intermediate phase angles, with the highest detection fractions near $60$--$70^\circ$.

\item The intermediate-phase optimum results from competing atmospheric and observational effects: molecular features become more favorable toward larger phases while planet brightness, angular separation, coronagraph accessibility, and achievable characterization SNR limit successful observations.

\item The new \texttt{EXOSIMS} framework propagates the phase-dependent atmospheric requirement directly into the characterization integration-time calculation and records whether the achieved observation satisfies the molecular detection criterion.

\item This enables a species-specific molecular detection yield to be calculated across complete simulated missions, connecting planet detection and atmospheric characterization to a mission-level atmospheric science metric.

\item Relative to fixed characterization thresholds of SNR = 5 for H$_2$O and SNR = 8.5 for O$_2$, the phase-dependent treatment increases the mean molecular detection yield by 76.8 percent for H$_2$O and 29.7 percent for O$_2$.

\end{itemize}

\section{Acknowledgments} \label{sec:acknowledgments}
SF gratefully acknowledges support from the Jet Propulsion Laboratory Strategic University Research Partnerships (SURP) Program and the University of Arizona through the State of Arizona's Technology and Research Initiative Fund. SF would also like to thank her HABLab group members and collaborators at NASA JPL for their support. TDR appreciates funding support from both NASA's Exoplanets Research Program (No. 80NSSC25K7149) and  NASA's Nexus for Exoplanet System Science Virtual Planetary Laboratory (No.~80NSSC23K1398). Part of this research was carried out at the Jet Propulsion Laboratory, California Institute of Technology, under a contract with the National Aeronautics and Space Administration (80NM0018D0004).

\clearpage


\bibliographystyle{plainnat} 
\bibliography{Footeexport} 

\end{document}